\documentclass[aps,prl,superscriptaddress,twocolumn,longbibliography]{revtex4-2}
\usepackage{bbm}
\usepackage{graphicx}
\usepackage{dcolumn}
\usepackage{bm}
\usepackage{subfigure}
\usepackage{amsmath}
\usepackage{amssymb}
\usepackage{feynmf}
\usepackage{hyperref}
\usepackage{color}

\usepackage{attachfile}

\usepackage{times}

\begin{document}

\title{Topological Hall Effect Driven by Short-Range Magnetic Orders in EuZn$_2$As$_2$}

\author{ Enkui Yi}
\affiliation{Center for Neutron Science and Technology, Guangdong Provincial Key Laboratory of Magnetoelectric Physics and Devices, School of Physics, Sun Yat-Sen University, Guangzhou, Guangdong, 510275, China}
\affiliation{State Key Laboratory of Optoelectronic Materials and Technologies, Sun Yat-Sen University, Guangzhou, Guangdong 510275, China}

\author{Dong Feng Zheng}
\affiliation{Songshan Lake Materials Laboratory Dongguan, Guangdong, 523808, China}

\author{Feihao Pan}
\affiliation{Laboratory for Neutron Scattering and Beijing Key Laboratory of Optoelectronic Functional Materials and MicroNano Devices, Department of Physics, Renmin University of China, Beijing 100872, China}

\author{Hongxia Zhang}
\affiliation{Laboratory for Neutron Scattering and Beijing Key Laboratory of Optoelectronic Functional Materials and MicroNano Devices, Department of Physics, Renmin University of China, Beijing 100872, China}

\author{Bin Wang}
\affiliation{Center for Neutron Science and Technology, Guangdong Provincial Key Laboratory of Magnetoelectric Physics and Devices, School of Physics, Sun Yat-Sen University, Guangzhou, Guangdong, 510275, China}
\affiliation{State Key Laboratory of Optoelectronic Materials and Technologies, Sun Yat-Sen University, Guangzhou, Guangdong 510275, China}

\author{Bowen Chen}
\affiliation{Center for Neutron Science and Technology, Guangdong Provincial Key Laboratory of Magnetoelectric Physics and Devices, School of Physics, Sun Yat-Sen University, Guangzhou, Guangdong, 510275, China}
\affiliation{State Key Laboratory of Optoelectronic Materials and Technologies, Sun Yat-Sen University, Guangzhou, Guangdong 510275, China}

\author{Detong Wu}
\affiliation{Center for Neutron Science and Technology, Guangdong Provincial Key Laboratory of Magnetoelectric Physics and Devices, School of Physics, Sun Yat-Sen University, Guangzhou, Guangdong, 510275, China}
\affiliation{State Key Laboratory of Optoelectronic Materials and Technologies, Sun Yat-Sen University, Guangzhou, Guangdong 510275, China}

\author{Huili Liang}
\affiliation{Songshan Lake Materials Laboratory Dongguan, Guangdong, 523808, China}

\author{Zeng Xia Mei}
\affiliation{Songshan Lake Materials Laboratory Dongguan, Guangdong, 523808, China}

\author{Hao Wu}
\affiliation{Songshan Lake Materials Laboratory Dongguan, Guangdong, 523808, China}

\author{Shengyuan A. Yang}
\affiliation{Research Laboratory for Quantum Materials, Singapore University of Technology and Design, Singapore 487372, Singapore}

\author{Peng Cheng}
\email{pcheng@ruc.edu.cn}
\affiliation{Laboratory for Neutron Scattering and Beijing Key Laboratory of Optoelectronic Functional Materials and MicroNano Devices, Department of Physics, Renmin University of China, Beijing 100872, China}

\author{Meng Wang}
\email{wangmeng5@mail.sysu.edu.cn}
\affiliation{Center for Neutron Science and Technology, Guangdong Provincial Key Laboratory of Magnetoelectric Physics and Devices, School of Physics, Sun Yat-Sen University, Guangzhou, Guangdong, 510275, China}

\author{Bing Shen }
\affiliation{Center for Neutron Science and Technology, Guangdong Provincial Key Laboratory of Magnetoelectric Physics and Devices, School of Physics, Sun Yat-Sen University, Guangzhou, Guangdong, 510275, China}
\affiliation{State Key Laboratory of Optoelectronic Materials and Technologies, Sun Yat-Sen University, Guangzhou, Guangdong 510275, China}

\date{\today}

\begin{abstract}
Short-range (SR) magnetic orders such as magnetic glass orders or fluctuations in a quantum system usually host exotic states or critical behaviors. As the long-range (LR) magnetic orders, SR magnetic orders can also break time-reversal symmetry and drive the non-zero Berry curvature leading to  novel transport properties. In this work, we report that in EuZn$_2$As$_2$ compound, besides the LR A-type antiferromagnetic (AF) order, the SR magnetic order is observed in a wide temperature region. The magnetization measurements and electron spin resonance (ESR) measurements reveal the ferromagnetic (FM) correlations for this SR magnetic order which results in an obvious anomalous Hall effect above the AF transition. Moreover the ESR results reveal that this FM SR order coexists with LR AF order exhibiting anisotropic magnetic correlations below the AF transition. The interactions of LR and SR magnetism evolving with temperature and field can host non-zero spin charility and berry curvature
leading the additional topological Hall contribution even in a centrosymmetric simple AF system. Our results indicate that EuZn$_2$As$_2$ is a fertile platform to investigate exotic magnetic and electronic states.

\end{abstract}
\maketitle

Short-range (SR) magnetic orders are widely investigated and draw great interest in various quantum systems in condensed matter physics. Compared to long-range (LR) magnetic orders, SR magnetic orders such as magnetic glass orders (static) or  magnetic fluctuations (dynamic) etc can emerge at the temperatures much higher than those for the establishment of LR magnetic orders, accompanied with exotic orders such as unconventional superconducting order, electron nematic order etc hosting the critical behaviors and novel quantum phenomena \cite{Kivelson, von}. For a topological system, the LR magnetic orders are usually considered to develop magnetic nontrivial topological states due to robust magnetism and large magnetic gaps. Recently, the observed SR magnetic orders in topological materials such as EuCd$_2$As$_2$ or MnBi$_2$Te$_4$ also drive magnetic nontrivial topological states and host the abnormal transport behaviors above the antiferromagnetic (AF) transition due to time-reversal symmetry breaking \cite{Rahn,Majz,Caoxy,Xuy,Alfonsov}. This enlightens us to consider SR magnetic orders such as magnetic fluctuations, which may be underlooked for a long time in topological systems, to design the magnetic nontrivial topological states or achieve the topological effects in the high-temperature (or even room-temperature) region \cite{GeJ,Ghimire}.

In Eu-122 system, such as EuCd$_2$As$_2$, EuSn$_2$As$_2$ or EuIn$_2$As$_2$, besides the AF transition at low temperatures ($T_{AF}$),  SR magnetic orders were observed in a wide temperature region \cite{Rahn,Majz,Jonh,Taddei,Zhangy,Yiek,Riber,Sunhl}. For examples, in EuCd$_2$As$_2$, strong spin fluctuations emerge  around 100 K and drive the magnetic Weyl Fermions far above the AF transition \cite{Majz}. Correspondingly, unconventional anomalous Hall and Nernst effects also emerge above $T_{AF}$ and behave different temperature and field evolvement from those below $T_{AF}$ \cite{Xuy}. In EuIn$_2$As$_2$ (considered as an axion insulator), magnetic polaron (MP) is revealed above $T_{AF}$ leading to large negative magneto-resistivity (MR) \cite{Zhangy}. In addition, helical magnetic orders accompanied with $A$-type AF lattice below $T_{AF}$ were also observed in some research indicating complicate magnetic structure for EuIn$_2$As$_2$ \cite{Riber}. Eu-122 compounds provide a fertile playground for investigating the interplay between various magnetism and electron topology hosting novel transport properties. Different from metallic EuCd$_2$As$_2$ or EuIn$_2$As$_2$, EuZn$_2$As$_2$ is a semiconducting compound but shares the similar crystal structure. Although the simple AF transition is observed, the transport features above this transition may suggest the presence of complicated magnetic orders and interactions which are still far beyond well understood\cite{Blawat,Bukow,Wangzc}. In this work, we perform a systematic study for semiconducting EuZn$_2$As$_2$.  An in-plane AF structure is identified by our single-crystal neutron scattering measurements around $T_{AF}$ = 20 K. Besides this LR magnetic order, the SR ferromagnetic (FM) order is revealed by electron spin resonance (ESR) and magnetization measurements. This SR FM order spans over a wide temperature region even above $T_{AF}$ and drives the anomalous Hall effect (AHE).  More interestingly, the prominent topological Hall effect (THE) is observed in a centrosymmetric system in absence of helical magnetic orders. With decreasing temperature to 12 K, the THE exhibits a small shoulder due to the change of  the interaction between LR AF and SR FM orders. These abnormal behaviors suggest the complicated interaction of LR and SR in EuZn$_2$As$_2$ can drive large Berry curvature and non-zero spin chirality hosting multiple novel transport properties.

\begin{figure}
  \centering
  \includegraphics[width=3in]{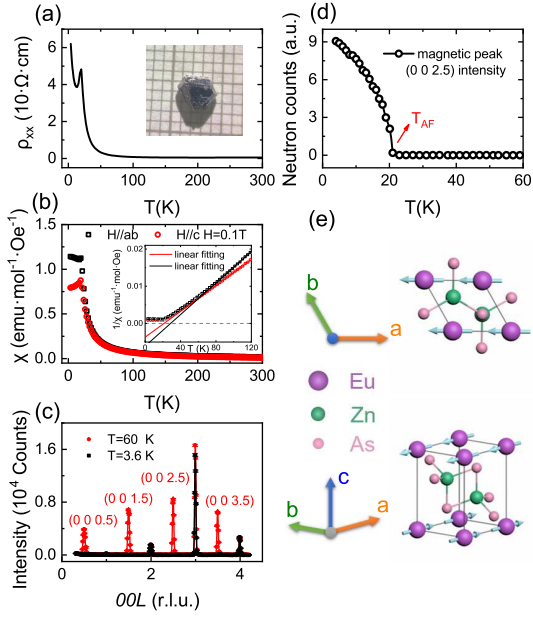}
  \caption{(a) Temperature dependent longitudinal resistivity $\rho_{xx}$($T$). (b) Temperature dependence of magnetic susceptibilities ($\chi$($T$)) with the applied field of 0.1 T  perpendicular ($\mu_0H//ab$) and parallel ($\mu_0H//c$) to the $c$ axis of the crystal respectively. The inset: the inverses of magnetic susceptibilities for $\mu_0H//ab$ and parallel $\mu_0H//c$ fitted by Curie–Weiss law. (c) Neutron diffraction scans along the (00L) reciprocal-lattice direction for 3.6 K and 60 K. (d) The neutron intensity of magnetic peak (0 0 2.5) as a function of temperature. (e) The schematic crystal and magnetic structure of EuZn$_2$As$_2$ below $T_{AF}$.
}
  \label{fig:Fig1}
\end{figure}

Single crystals of EuZn$_2$As$_2$ were grown by the Sn-flux method \cite{Bukow,Blawat,Wangzc}.  Structure and elemental composition of crystals were assessed using X-ray diffraction and energy dispersive X-ray spectroscopy respectively. Single-crystal neutron diffraction experiment was carried out on Xingzhi cold neutron triple-axis spectrometer at the China advanced research reactor (CARR) \cite{ChengP}. A single crystal with the mass of 0.1g was aligned to the (HHL) scattering plane. The incident neutron energy was fixed at 15 meV with a neutron velocity selector used upstream to remove higher-order neutrons. Transport measurements were performed on a commercial physical property measurement system (PPMS Dynacool, Quantum Design). Magnetization measurements were carried on a vibrating sample magnetometer (VSM) based on the PPMS. ESR signals were obtained by a Bruker EMX plus X-band  (9.365 GHz) CWEPR spectrometer.

As shown in Fig. 1(a), EuZn$_2$As$_2$ exhibits semiconducting behavior revealed by the temperature dependent resistivity ($\rho_{xx}$($T$)) in sharp contrast to metallic EuCd$_2$As$_2$ or EuIn$_2$As$_2$ \cite{Niucw,Sohjr,Zhangy}. An abnormal peak was observed around 20 K both in $\rho_{xx}$($T$)  and temperature dependent magnetization ($M$($T$)) due to a magnetic transition shown in Fig. 1(b). To check this transition and related magnetic order, single-crystal neutron diffraction measurements were performed and the scans along the (00L) reciprocal-lattice direction are shown in Fig. 1(c). The magnetic Bragg peaks at L = 0.5n (n = odd number) are observed at 3.5 K but absent at 60 K. Besides, the magnetic Bragg peaks indexed by (11L) with L = 0.5n (n = odd number) were also identified.  After carefully checking diffraction results for several other directions, we did not observed additional incommensurate magnetic Bragg peaks as those in EuIn$_2$As$_2$ \cite{Riber}. The neutron diffraction results support an $A$-type spin configuration with moments lying in the ab-plane below transition temperature $T_{AF}$ without appearance of complex helical magnetic structure consistent with previous report \cite{Riber} shown in Fig. 1(e). The temperature dependent intensity of magnetic Bragg peak (0,0,2.5) is shown in Fig. 1(d). The peak intensity quickly drops to background value at a transition temperature ($T_{AF}$) indicating the long-range antiferromagnetic order varnishing above $T_{AF}$ in consistent with the magnetization measurements results.

\begin{figure}
  \centering
  \includegraphics[width=3in]{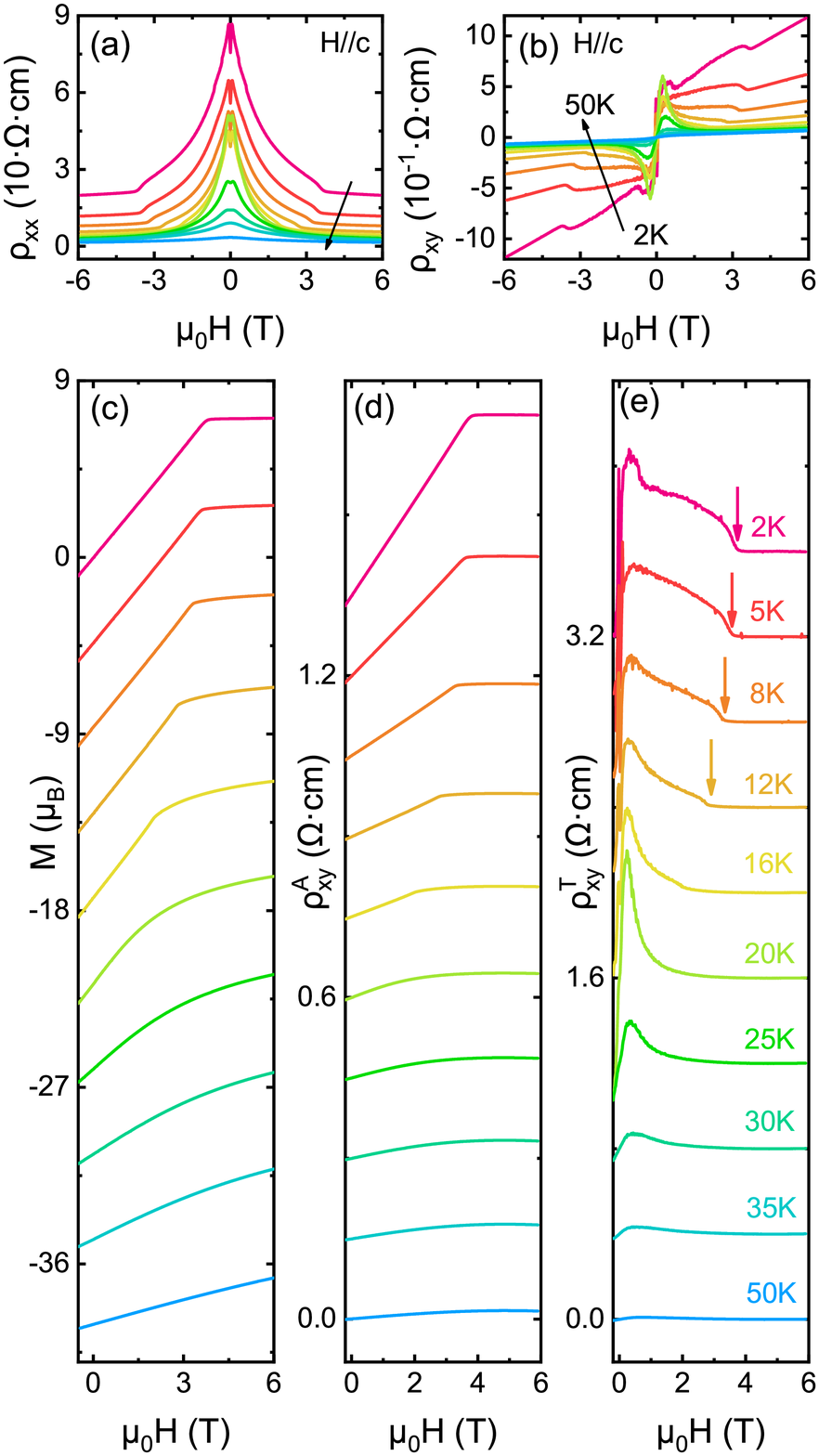}
  \caption{ The field dependent longitudinal resistivity $ \rho_{xx}$($\mu_0H$) (a), and Hall resistivity $\rho_{xy}$($\mu_0H$) (b) at 2, 5, 8, 12, 16, 20, 25, 30, 35, 50 K respectively. According to the formula $\rho_{xy}=\rho^N_{xy}+\rho^A_{xy}+\rho^T_{xy}=R_0\mu_0H+AM+\rho^T_{xy}$, the AHE component $\rho^{A}_{xy}$ and THE component $\rho^T_{xy}$ are extracted from the total Hall resistivity $\rho_{xy}$. The field dependent magnetization $M$($\mu_0H$) (c), anomalous Hall resistivity $\rho^{A}_{xy}$($\mu_0H$) (d), and topological Hall resistivity $\rho^T_{xy}$($\mu_0H$) (e)  at 2, 5, 8, 12, 16, 20, 25, 30, 35, 50 K respectively. The absolute values of $M$($\mu_0H$), $\rho^{A}_{xy}$($\mu_0H$), and $\rho^T_{xy}$($\mu_0H$) shift at various temperatures.  In $\rho^T_{xy}$($\mu_0H$) curves, the second-shoulder features are marked by arrows.
 }
  \label{fig:Fig2}
\end{figure}

The field dependent longitudinal resistivity ($\rho_{xx}$($\mu_0H$)) and Hall resistivity ($\rho_{xy}$($\mu_0H$)) exhibit systematical temperature evolving as shown in Figs. 3(a) and (b). Prominent negative MR is observed at low temperatures. With increasing the temperature, this prominent negative MR becomes weak gradually but can persist up to the temperature (around 100 K) far above $T_{AF}$. Correspondingly, $\rho_{xy}$($\mu_0H$) exhibits nonlinear field dependence below 100 K indicating the presence of an extra Hall contribution (AHE) from the magnetism or non-zero Berry curvature. With further decreasing the temperature (below 30 K), after subtracting the normal Hall contribution due to the Lorentz force with the linear field response, a discrepancy between $\rho_{xy}$($\mu_0H$) and $M$($\mu_0H$) is observed suggesting the emergence of an additional Hall contribution (THE) besides AHE. Thus, the total Hall resistivity for EuZn$_2$As$_2$ can be expressed as: \cite{Nagaosa,WangB}
\begin{equation}
 \rho_{xy}=\rho_{xy}^N+\rho_{xy}^A+\rho_{xy}^T=R_0 \mu_0H+AM+\rho_{xy}^T
\end{equation}
where $\rho_{xy}^N$, $\rho_{xy}^A$, and $\rho_{xy}^T$ are the normal Hall resistivity, anomalous Hall resistivity and topological Hall resistivity respectively, while $R_0$ and $A$ are the Hall coefficient and the anomalous Hall coefficient. According to this formula and the data of $M$($\mu_0H$), $\rho_{xy}^A$ and $\rho_{xy}^T$ are separated from $\rho_{xy}$  and presented in Figs. 2(c)-(d).

The first interesting observation is the prominent AHE spanning over a large temperature region even with the absence of AF order. For an AF system, it is usually surprising to host obvious AHE due to the absence of net magnetization. But the recent experimental researches revealed the chiral magnetic structure or Berry curvature could also host the large AHE based on a AF structure \cite{Nakatsuji,Liuc,Dengyj}. For EuZn$_2$As$_2$, only a simple colinear AF order is identified by our measurements in contrast to EuCd$_2$As$_2$ or EuIn$_2$As$_2$ which hosts an additional helical magnetic order besides the basic colinear AF structure \cite{Riber}. Such simple magnetic structure in EuZn$_2$As$_2$ seems too difficult to host large net magnetization indicating a possible different origin for the large AHE.  Another observed abnormal transport feature is the prominent THE. At low temperatures, it exhibits two-peak feature (a peak around zero field with a small shoulder in larger field region marked by arrows in Fig. 2(e)). With increasing the temperature, the small shoulder become weak gradually and invisible at 12 K eventually. In addition, THE can also persist to the temperature region above $T_{AF}$ where the AF LR magnetic order vanishes. Usually THE is considered to originate from the movement of skyrmions in noncentrosymmertric non-colinear magnets hosting the nonzero scalar spin chirality \cite{Nagaosa}. However, these physics pictures can not describe the case of EuZn$_2$As$_2$ with a simple colinear AF magnetic structure and a centrosymmertric crystal structure.

\begin{figure}
  \centering
  \includegraphics[width=3in]{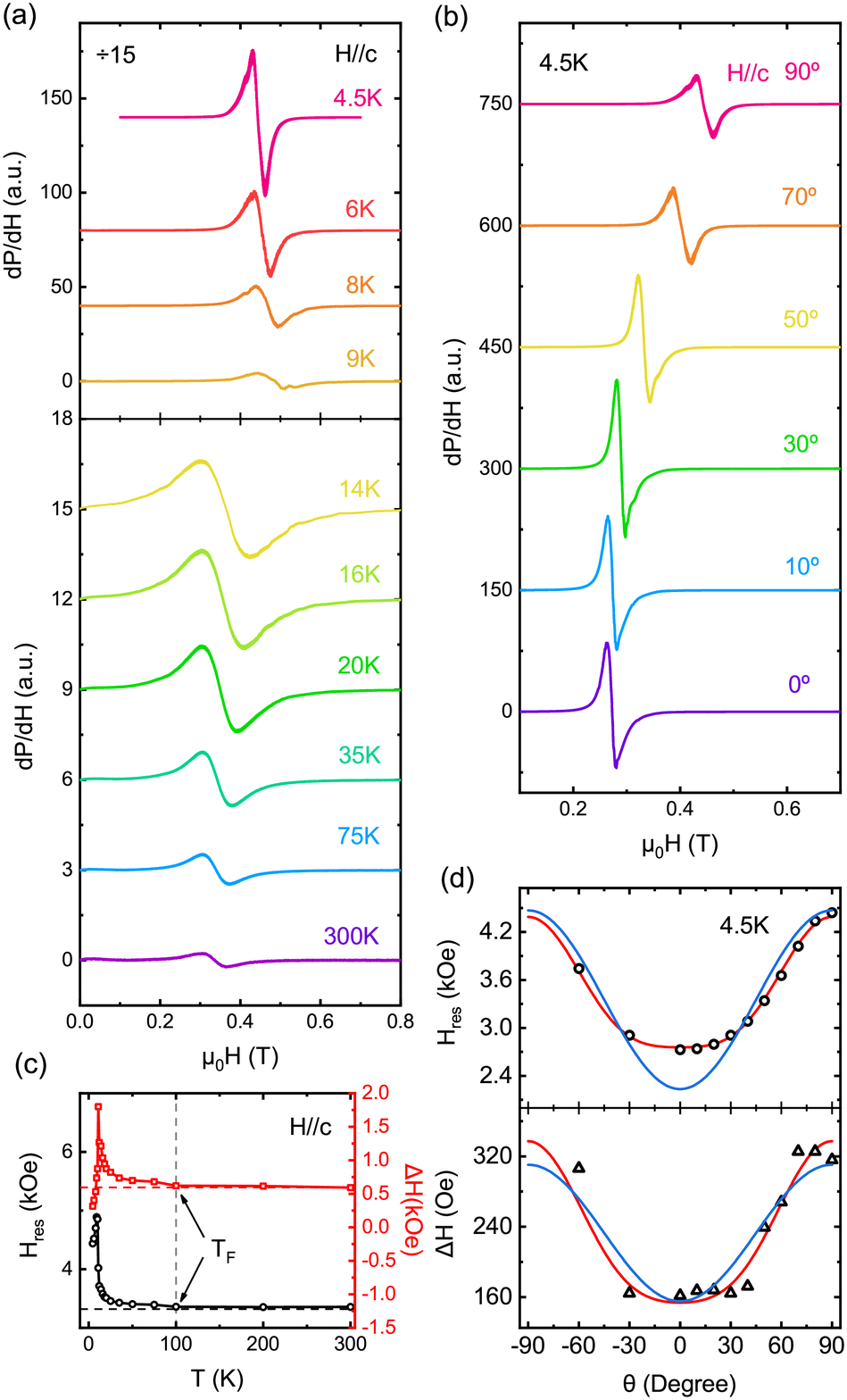}
  \caption{(a) ESR spectra of EuZn$_2$As$_2$ at 4.5, 6, 8, 9, 14, 16, 20, 35, 75, and 300 K with the applied field along the $c$ axis. (b) ESR spectra of EuZn$_2$As$_2$ with $\theta $ from  0$^{\circ}$ to 90$^{\circ}$, where $\theta$ is the angle between the applied field and the $ab$ plane of the crystal. (c) Temperature dependence of  resonance field $H_{res}$($\theta$) (left) and line width $\Delta H$ (right). (d) Angular dependence of $H_{res}$ ($\theta$) (upper) and $\Delta H$($\theta$) (lower) at 4.5K. Red lines represent the  fitting curves for $H_{res}$($\theta$) and $\Delta H$($\theta$) by formulas as $H_{res}$ = $a_1$(1+sin$^2$($\theta$)+$a_2$(3cos$^2$($\theta$)-1)$^2$) and $\Delta H$ = $a_1$(1+sin$^2$($\theta$)+$a_2$(3sin$^2$($\theta$)-1)$^2$) respectively, where $a_1$ and $a_2$ are fitting coefficients. Blue line can be expressed as $H_{res}$ = $a_1$(1+sin$^2$($\theta$)) and $\Delta H$ = $a_1$(1+sin$^2$($\theta$))  \cite{Zeisner}.
 }
  \label{fig:Fig3}
\end{figure}

To understand these AHE and THE, short-range magnetic order, which may be underlooked for a topological protected system, need to be considered in EuZn$_2$As$_2$. It is observed that the $M$($\mu_0H$) curves for both $\mu_0H//ab$ and $\mu_0H//c$ start to deviate the Curie-Weiss behavior at a temperature $T_F$ (around 100 K) much higher than $T_{AF}$ coincided with presence of negative MR. These abnormal transport behaviors reveal large-scale SR magnetic interactions before the establishment of the LR AF orders. To further investigate this SR order, ESR measurements were performed in a wide temperature region as shown in Fig. 3. All ESR spectra ((d$P$/d$H$) vs $H$)  exhibit a single exchange-narrowed resonance without other hyperfine lines as shown in Figs. 3(a) and (b). At high temperatures, the ESR signals exhibit an asymmetric Dysonian shape which is characteristic of localized magnetic moments in a lattice with a skin depth for a single crystal \cite{Alfonsov,Feher,Dyson}. With decreasing the temperature, the ESR signals become stronger and deviate from the Dysonian-shape relation gradually. It is observed the ESR line width $\Delta H$ exhibits weak temperature dependence above $T_F$ (100 K) in consistent with the resonant field $H_{res}$ keeping 3350 Oe associated with the g-factor ($g = h\nu/\mu_0H_{res}$, where $\nu$ is electromagnetic wave frequency) of 1.99 in the same temperature region  where  localized Eu$^{2+}$ 4$f$ electron spins dominate this paramagnetic state \cite{Gory,Goryunov}. Below $T_F$, $\Delta H$ starts to increase while the $H_{res}$  decreases with decreasing the temperature above $T_{AF}$.  These behaviors reveal that an effective internal magnetic field develops as the magnetic correlations  with cooling the system down to $T_F$.  And it is noticed that $T_{F}$ is roughly five times larger than $T_{AF}$ which indicates  magnetic interactions drive SR magnetic order over a large temperature scale above $T_{AF}$.

Below $T_{AF}$ for $H // ab$, $H_{res}$  increases to the value of 2728Oe at 4.5 K associated with the increase of $M$($T$) during cooling process which is in consistent with positive Curie-Weiss temperature of $\Theta$ = 25 K acquired by fitting $M$($T$) curves. These behaviors reveal a FM correlation for in-plane magnetic interactions favoring the $A$-type AF structure revealed by our neutron results. In contrast, for $H // c$ $H_{res}$  drops sharply with decreasing temperature below $T_{AF}$ and to 1510 at 4.5 K in consistent with the decrease of $M$($T$) suggesting the AFM correlation for inter-plane magnetic interactions. But the $\Theta$ = 20 K acquired by fitting $M$($T$) curves for $H // c$ reveals that the FM correlation still persists accompanied with an $A$-type AF order.
As shown in Fig. 3(d), the angular ESR line width $\Delta H$($\theta$) and resonant field $H_{res}$($\theta$) (where $\theta$ is the angle between the applied field and the $ab$ plane)  reveal the aniostropic magnetic correlations. In a weak correlated spin system such as paramagnetic(PM) or weak correlated AF system, $H_{res}$($\theta$) and $\Delta H$($\theta$) follow the relations of $H_{res}$($\theta$) $\sim$ (sin$^2$($\theta$) + 1) and $\Delta H$($\theta$) $\sim$ (cos$^2$($\theta$) + 1) respectively. These angular dependent laws can well describe the anisotropy due to the interactions for uncorrelated spins such as that for the type-A AF state in MnBi$_2$Te$_4$ or for the PM state in intrinsically low-dimensional van der Waals magnets\cite{Alfonsov,Zeisner}. Here our angular data violate these relations suggesting stronger and more anisotropic magnetic correlations. It is observed our $\Delta H$($\theta$) and $H_{res}$($\theta$) follow
the relations of $H_{res}$ = $a_1$(1+sin$^2$($\theta$)+$a_2$(3cos$^2$($\theta$)-1)$^2$) and $\Delta H$ = $a_1$(1+sin$^2$($\theta$)+$a_2$(3sin$^2$($\theta$)-1)$^2$) respectively (where $a_1$ and $a_2$ are fitting coefficients) which describe the typical two-dimensional magnetic systems with the increasing dominance of long-wavelength fluctuations such as  the  FM state in Cr$_2$Ge$_2$Te$_6$\cite{Zeisner}. These anisotropic magnetization and ESR results suggest: (1) For the EuZn$_2$As$_2$ with layered crystal structure, the in-plane FM interactions are expected to be much stronger than the interlayer AFM interactions. (2) The FM correlation spans over a large temperature scale for both in-plane and out-of-plane directions \cite{Majz}. (3) The substantial frustrations which may be due to nearest- and next nearest–neighbor magnetic exchange couplings drive an anisotropic SR FM order strongly interacting with the static LR AFM order.

\begin{figure}
  \centering
  \includegraphics[width=3in]{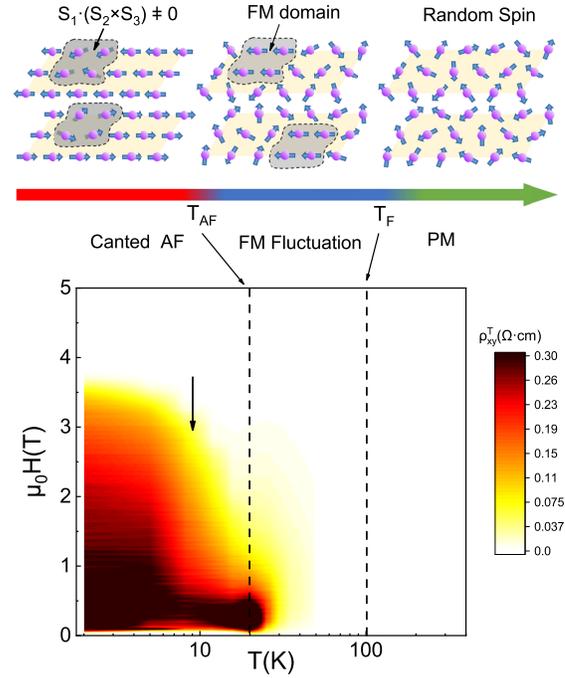}
  \caption{The phase diagram for EuZn$_2$As$_2$. The upper: The schematic of various magnetic states in different temperature regions. The canted AF state can host local non-zero spin chirality. FM fluctuations can host local FM domain. For paramagnetic state(PM) the spins are random and isotropic. The lower:  The color plot for topological Hall resistivity. The black broken lines divide the diagram into three regions. The arrow around 12 K indicates the vanishing of second shoulder for THE.}
  \label{fig:Fig4}
\end{figure}

Fig. 4 shows the phase diagram of EuZn$_2$As$_2$. In a frustrated system, with cooling the system from the PM state, the in-plane FM correlations increase and the dipolar interactions will then further stabilize FM correlations out of the plane. The related SR FM order divides the system into local FM domains whose net-magnetization brings the additional AHE term for $\rho_{xy}$ below $T_{F}$. With further decreasing the temperature, both dipolar and AFM interlayer exchange interactions prefer magnetization changes from an out-of-plane to an in-plane orientation leading to an $A$-type AFM structure. However, the FM SR order still persists and interacts with the established long-range order leading to a canted AF structure. These canted spins with local FM orders can also host non-zero spin chirality in a centrosymmertric system as that in EuCd$_2$As$_2$ leading to the additional THE contribution which can even persist above $T_{AF}$ \cite{Wangll,Sohjr,Xuy}. For a time, the SR magnetic orders were underlooked for developing a robust magnetic topological system. But the recent studies in Mn-Bi-Te and Eu-122 systems revealed that the SR magnetic orders \cite{LeeSH} such as magnetic fluctuations or magnetic polarons can also change the topological nature for a magnetic system as the SR magnetic orders which may drive novel feature at higher temperatures such as the observed magnetic Weyl point far above $T_{AF}$ \cite{Alfonsov,Majz,Zhangy}.  For our EuZn$_2$As$_2$, when the magnetic correlations become strong enough, the system can also change the local magnetic chirality and Berry curvature probably mediated by the magnetic fluctuations hosting the novel transport properties. Moreover, it is observed that THE exhibits second-shoulder feature below 12 K which may suggest a crossover by the competing interaction of the LR AF order and the SR FM order. In EuZn$_2$As$_2$, due to the evolved magnetic correlations the SR magnetic order can persist and interact with electron and LR magnetic order in a wide scale resulting in various abnormal transport features.

In summary, we systemically investigate the magnetism and the related transport properties for EuZn$_2$As$_2$. The SR FM order is revealed in a prominent temperature region above the AF transition. This short-range order results in a large AHE above $T_{AF}$ and canted AF orders below $T_{AF}$. These interacted magnetic orders drive THE in a centrosymmetric system. Our results indicate the short-range magnetic order can bring the non-zero spin charility and the Berry curvature driving novel transport properties which should be also considered and emphasized for the investigation of the interplay between magnetism and topology in quantum materials.

\section{Acknowledgements} Work are supported by National Natural Science Foundation of Chin (NSFC) (Grants No.U213010013, 92165204, 11904414, 12174454, and 11904417),  Natural Science Foundation of Guangdong Province (Grant No. 2022A1515010035,2021B1515020026), open research fund of Songshan Lake materials Laboratory 2021SLABFN11, National Key Research and Development Program of China (Grant No. 2019YFA0705702), OEMT-2021-PZ-02, and Physical Research Platform (PRP) in School of Physics, Sun Yat-sen University.



%

\end{document}